\documentclass[twocolumn,showpacs,preprintnumbers,amsmath,amssymb,A4paper]{revtex4}

\usepackage[dvips]{graphicx}
\usepackage{dcolumn}
\usepackage{bm}

\begin{document}
\newcommand{\kp}{{\bf k$\cdot$p}\ }

\preprint{APS/123-QED}
 \title{Temperature dependence of the electron spin $\textbf{g}$ factor in GaAs}

\author{W. Zawadzki$^*$, P. Pfeffer}
 \affiliation{Institute of Physics, Polish Academy of Sciences\\
 Al.Lotnikow 32/46, 02--668 Warsaw, Poland\footnotetext{$^*$ e-mail address: zawad@ifpan.edu.pl}\\}

\author{R. Bratschitsch$^{**}$, Z. Chen, S. T. Cundiff}
\affiliation{JILA, National Institute of Standards and Technology
and University of Colorado, Boulder, Colorado 80309-0440,
USA\\\footnotetext{ $^{**}$ Present address: University of
Konstanz and Center for Applied Photonics, D-78457 Konstanz,
Germany}}

\author{B. N. Murdin}
 \affiliation{Advanced Technology Institute, University of Surrey, Guildford GU2 7XH, United Kingdom\\}

\author{C. R. Pidgeon}
 \affiliation{Department of Physics, Heriot-Watt University, Edinburgh EH14 4AS, United Kingdom\\}

\date{\today}

\begin{abstract}
The temperature dependence of the electron spin $g$ factor in GaAs
is investigated experimentally and theoretically. Experimentally,
the $g$ factor was measured using time-resolved Faraday rotation
due to Larmor precession of electron spins in the temperature
range between 4.5 K and 190 K. The experiment shows an almost
linear increase of the $g$ value with the temperature. This result
is in good agreement with other measurements based on
photoluminescence quantum beats and time-resolved Kerr rotation up
to room temperature. The experimental data are described
theoretically taking into account a diminishing fundamental energy
gap in GaAs due to lattice thermal dilatation and nonparabolicity
of the conduction band calculated using a five-level {\kp} model.
At higher temperatures electrons populate higher Landau levels and
the average $g$ factor is obtained from a summation over many
levels. A very good description of the experimental data is
obtained indicating that the observed increase of the spin $g$
factor with the temperature is predominantly due to band's
nonparabolicity.

\end{abstract}

\maketitle

\section{\label{sec:level1}Introduction\protect\\ \lowercase{}}

The temperature dependence of the spin $g$ value of electrons in GaAs has been a subject of controversy since
1995, when it was shown that the experimental data, exhibiting an increase of the value of $g$ as a function of
temperature, are in a qualitative disagreement with the {\kp} theory, if one takes into account an experimental
change of the fundamental energy gap [1]. The seeming qualitative disagreement between the experiment and the
theory was confirmed by subsequent publications [2, 3]. The temperature dependence of the spin properties of
electrons in GaAs is not only of academic interest since for possible spintronic applications the behavior of
$g$ near room temperature is clearly of great importance. It was recently shown that one can reach at least a
qualitatively correct description of the experimental $g$ values if one includes in the {\kp} theory not the
complete temperature change of the fundamental energy gap but the change due to lattice dilatation alone [4]. A
similar result was obtained for the $g$ value of electrons in InSb indicating that the better description was
not fortuitous.

The experiments [1-3] and their analysis presented in Ref. [4]
have two shortcomings. On the experimental side, the $g$ value
data was obtained by time-resolved photoluminescence alone, while
it would be desirable to verify these data by some other method.
On the theoretical side, the analysis in Ref. [4] is somewhat
tentative as it does not account for the fact that at higher
temperatures electrons populate many Landau levels, so that the
measured spin $g$ value represents an average over many
transitions. Recognizing the importance of the subject and its
controversial character the present work tries to overcome the
above shortcomings in both experimental and theoretical aspects.
First, we present new data on the temperature dependence of the
electron $g$ value in GaAs obtained with the time-resolved Faraday
rotation and compare them with the photoluminescence data of Refs.
[1-3]. Second, we complete the theoretical description by
including the electron statistics which results in the necessity
of summation over many Landau levels (LLs). We show that the
complete theory considerably improves the description of
experimental data.

The paper is organized as follows. We first describe new experiments on the temperature dependence of the
electron $g$ value in GaAs. Next, we describe the theoretical procedure used to evaluate the temperature
dependence of the $g$ value and compare the theory with all existing experimental data. We discuss our results
and conclude the paper by a summary.

\section{\label{sec:level1}EXPERIMENT AND THEORY\protect\\ \lowercase{}}

The spin $g$ factor of electrons in a nominally undoped 30 $\mu$m thick layer of GaAs was determined using an
ultrafast degenerate setup for measuring the time-resolved Faraday rotation [5-7]. Spin polarized electrons were
optically excited by a circularly polarized pump pulse which was spectrally tuned to the band gap of GaAs at all
temperatures. After the excitation the electron spins begin to precess in a magnetic field applied parallel to
the sample surface in the Voigt geometry. By measuring the Faraday rotation of a linearly polarized probe beam
transmitted through the sample one is able to monitor the electron spin dynamics. The Larmor spin precession
results in an oscillating signal that decays exponentially at a rate 1/$T^*_2$, where $T^*_2$ is the ensemble
spin dephasing time (Fig. 1). The $g$ factor is determined using the equation of Larmor precession: $g^* =
\hbar\omega_L/\mu_BB$, where $\omega_L$ is the measured Larmor precession frequency, $\mu_B$ is the Bohr
magnetron, and $B$ is the applied magnetic field. The ultrafast femtosecond laser pulses ($\Delta\lambda$ = 15.5
nm, $P_{pump}$ = 10 mW, $P_{probe}$ = 1mW) were generated by a mode-locked Ti:sapphire laser oscillator
operating at 76 MHz repetition frequency. The samples were held at low temperatures using an optical cryostat
and a magnetic field was generated by a split-coil superconducting magnet. Magnetic fields up to 6 T have been
applied.

\begin{figure}
\includegraphics[scale=0.95,angle=0, bb = 20 25 282 220]{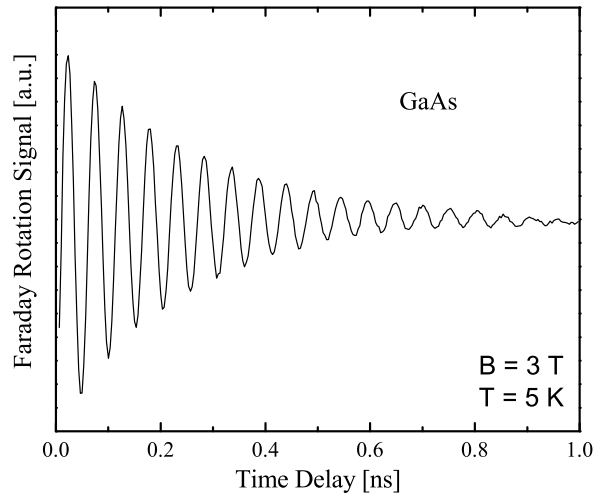}
\caption{\label{fig:epsart}{Time-resolved Faraday rotation signal of nominally undoped bulk GaAs, recorded at a
magnetic field of $B$ = 3 T.}} \label{fig1th}
\end{figure}

GaAs is a medium-gap material, so that a three-level {\kp}
description, successfully used for narrow gap semiconductors [8,
9], is not adequate for treating its band structure. The reason is
that in GaAs the fundamental gap $E_0$ between the $\Gamma^c_6$
and $\Gamma^v_8$ levels is about 1.5 eV, i.e. it is not much
smaller than the gap $E_1$ between the $\Gamma^c_6$ level and the
upper $\Gamma^c_7$ conduction level (which is about 3 eV). It has
been convincingly demonstrated that an adequate way to treat the
conduction band of GaAs is to use a five-level model (5LM, which
is equivalent to 14 bands including spin) in the {\kp} description
(see Refs. [10, 11] and the references therein). In particular,
for the description of electron spin $g$ value in GaAs-based
heterostructures the usefulness of the 5LM was demonstrated quite
recently [12]. Since the 5LM for electrons in the presence of a
magnetic field and its use for magnetooptical properties of GaAs
was described in some details before [10, 11], we only mention
here the main elements of this approach. Thus the model includes
exactly the $\Gamma^v_7$, $\Gamma^v_8$, $\Gamma^c_6$, $\Gamma^c_7$
and $\Gamma^c_8$ levels at the center of the Brillouin zone and
the resulting {\kp} matrix has dimensions 14 $\times$ 14. There
exist three nonvanishing interband matrix elements of momentum:
$P_0$, $P_1$ and $Q$, and one interband element of the spin-orbit
interaction $\overline{\Delta}$. If one takes Q = 0 and $k_z$ = 0
(where $k_z$ is the momentum along the magnetic field direction)
the 14 $\times$ 14 initial matrix factorizes into two 7 $\times$ 7
matrices for the spin-up and spin-down states. These matrices are
soluble with the envelope functions in the form of harmonic
oscillator functions and the eigenenergy problem for different
Landau levels (LLs) $n$ reduces to a diagonalization of 7 $\times$
7 determinants. However, if the $Q$ element is included (it comes
from an inversion asymmetry of the GaAs crystal) the initial 14
$\times$ 14 matrix does not factorize and it is not soluble in
terms of a single column of harmonic oscillator functions.
Physically, this means that the resulting energy bands are not
spherical. Since the nonsphericity of the conduction band in GaAs
is small, one can solve for the eigenenergies looking for the
envelope functions in terms of sums of harmonic oscillator
functions. This leads to number determinants composed of the
fundamental 7 $\times$ 7 blocks on the diagonal coupled by
nondiagonal parts involving the Q elements. The eigenenergies are
computed truncating the resulting big determinants. In our
computations we used typically 21 $\times$ 21 determinants. All
calculations were performed taking the magnetic field \textbf{
\emph{B}} parallel to [001] direction and putting $k_z$ = 0.

Now we turn to the temperature dependence of the fundamental gap
$E_0$. It was argued a long time ago [13] and confirmed by the
behavior of the effective mass [14, 15] that the temperature
dependence of the band parameters is governed by a dilatational
change of the energy gap E$_0$ and not by its total (i.e.
measured) change. The other contribution to the change of the gap
is due to phonons. However, the phonon vibrations occur on a much
slower time scale than the time the electron needs to sample the
interband interactions determining the effective mass and the $g$
factor, which are at optical frequencies. The dilatational change
of the gap is given by [16]
\begin{figure}
\includegraphics[scale=0.95,angle=0, bb = 20 25 282 220]{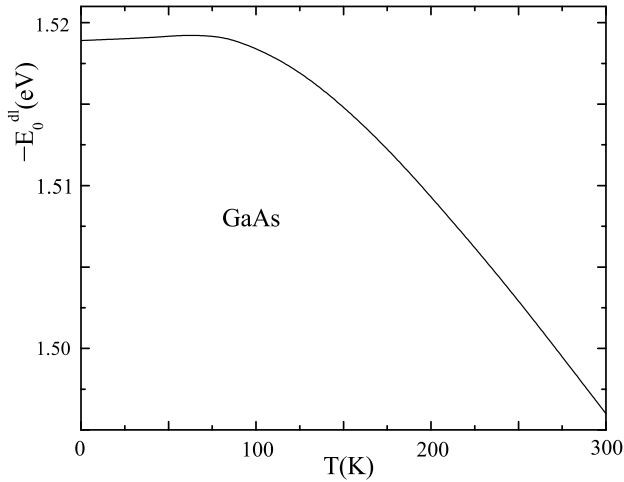}
\caption{\label{fig:epsart}{Calculated change of the fundamental energy gap in GaAs due to thermal dilatation
versus temperature (after [16]).}} \label{fig2th}
\end{figure}

\begin{equation}
\Delta E^{dl}_0(T)=-3D\left(\frac{\partial E}{\partial P}\right)_T \int^T_0 \alpha_{th}(T')dT' \;\;,
\end{equation}
where D is the bulk modulus, $dE/dP$ is the pressure-induced
band-gap shift, and $\alpha_{th}(T)$ is the linear thermal
expansion coefficient (see also [17]). The quantities $B$ and
$\partial E/\partial P$ are readily measurable, the quantity
$\alpha_{th}(T)$ was measured for GaAs by Novikova [18] and Soma
et al [19]. Using the values of $\alpha_{th}(T)$ one calculates
$\Delta E^{dl}_0(T)$. This was done by Lourenco et al [16], we
replot their results in Fig. 2. It should be mentioned that
$\alpha_{th}(T)$ goes through a minimum near $T\approx$ 50 K, this
results in a plateau and a flat maximum of E$^{dl}_0(T)$ seen in
Fig. 2. The dilatational change $\Delta E^{dl}_0$ of the
fundamental gap between 0 K and 300 K is about 23 meV (see Fig.
2), while the total change $\Delta E^{tot}_0$ is about 93 meV (cf.
Refs. [21, 22]). These numbers are important when one tries to
understand why putting the total change $\Delta E^{tot}_0(T)$ or
the dilatational change $\Delta E^{dl}_0(T)$ into the calculations
of $g$ value, one obtains very different results in the two cases.

Finally, we consider average values of the spin $g$ factor measured as a function of temperature. The
measurements are usually done in relatively pure samples having low free electron densities. The electrons are
excited across the gap into the conduction band and into both spin states. The spin states are equally
populated, but the circularly polarized light produces a well defined coherence between them. The excited
electrons quickly thermalize and are distributed among LLs according to the lattice temperature without losing
their spin or phase. Then they interfere and quantum beats in the photoluminescence or other effects are
observed from many LLs. According to this picture the observed signal represents an average over the populated
LLs in which the electron thermal distribution over LLs determines their contribution to the average $g$ value.
This explains why various experiments performed on weakly doped samples give basically the same results
depending only on the temperature.

We assume that the electron energies are
\begin{equation}
E_{nk_z}^{\pm}={\cal E}_n^{\pm}+\frac{\hbar^2k^2_z}{2m^*_0} \;\;,
\end{equation}
where $n$ is the LL number, $\pm$ signs correspond to the two spin
states, $k_z$ is the wavevector along the direction of
\textbf{\emph{B}}, and $m^*_0$ is the effective mass at the band
edge. The energies ${\cal E}_n^{\pm}$ contain the intricacies of
the band structure mentioned above, but we first assume a simple
parabolic dependence on $k_z$ to make the averaging tractable. The
spin g value is defined as (in formulas we use $g^*$ symbol)
\begin{equation}
g^*=(E^+_{nk_z}-E^-_{nk_z})/\mu_BB.
\end{equation}
An averaging procedure involves summation over $n$ and
integrations over $k_x$ and $k_z$. A simple calculation gives the
average value of $g^*$ in the form
\begin{figure}
\includegraphics[scale=0.94,angle=0, bb = 20 25 276 253]{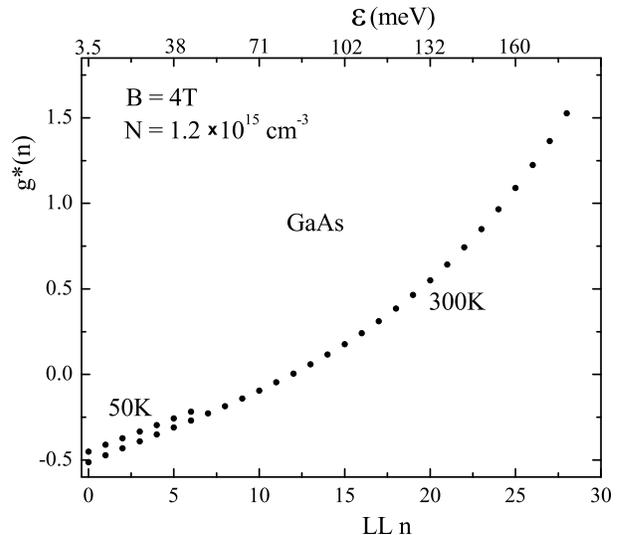}
\caption{\label{fig:epsart}{Spin $g$ values for consecutive Landau
levels $n$, calculated for $T$ = 50 K and $T$ = 300 K, a magnetic
field $B$ = 4 T and the indicated electron density N. The
indicated points correspond to LLs that give non-negligible
contributions to the average $g$ value. Approximate electron
energies corresponding to LLs marked on the lower abscissa are
shown on the upper abscissa.}} \label{fig3th}
\end{figure}

\begin{equation}
{\overline{g^*}(T)}=\frac{A}{C}\;\;,
\end{equation}
where
\begin{equation}
A=\sum^{\infty}_{n=0}\int^{\infty}_{{\cal E}^i_n} \frac{g^*_n({\cal E})f({\cal E,\zeta})}{({\cal E}-{\cal
E}^i_n)^{1/2}}d{\cal E}\;\;,
\end{equation}
and
\begin{equation}
C=\sum^{\infty}_{n=0}\int^{\infty}_{{\cal E}^i_n} \frac{f({\cal E,\zeta})}{({\cal E}-{\cal E}^i_n)^{1/2}}d{\cal
E} \;\;,
\end{equation}
in which the summation is over the LLs, $f({\cal E,\zeta})$ is the
Fermi-Dirac distribution function, and the square roots come from
the integrations over $k_z$. The integrations begin from the lower
states ${\cal E}^i_n$ for each $n$, which can be either ${\cal
E}^+_n$ or ${\cal E}^-_n$.

The average $g$ value, as given by Eq. (4), is affected by the temperature in two opposite ways. According to
the five-level model the spin $g$ value at the conduction band edge is [10],
$$
 g^*_0=2+\frac{2}{3}\left[E_{P_0}\left(\frac{1}{E_0}- \frac{1}{G_0}\right)+E_{P_1}\left(\frac{1}{G_1}-
 \frac{1}{E_1}\right)\right]+
$$
\begin{equation}
 -\frac{4{\overline{\Delta}}\sqrt{E_{P_0}E_{P_1}}}{9}\left(\frac{2}{{E_1}{G_0}}+
\frac{1}{{E_0}{G_1}}\right)+2C' \;\;,
\end{equation}
where $E_{P_0}$ = $2m_0P_0^2/\hbar^2$, $E_{P_1}$ =
$2m_0P_1^2/\hbar^2$, $G_0$ = $E_0 + \Delta_0$ and $G_1 = E_1 +
\Delta_1$. The spin-orbit energies $\Delta_0$ and $\Delta_1$
relate to ($\Gamma^v_7$, $\Gamma^v_8$) and ($\Gamma^c_7$,
$\Gamma^c_8$) levels, respectively, $\overline{\Delta}$ is the
interband matrix element of the spin orbit interaction between the
($\Gamma^v_7$, $\Gamma^v_8$) and ($\Gamma^c_7$, $\Gamma^c_8$)
multiplets (see [10, 22]), and $C'$ is due to far-band
contributions. For $\overline{\Delta}$ = 0 Eq. (7) reduces to the
formula given first by Hermann and Weisbuch [23].

It follows from Eq. (7) that, as the temperature $T$ increases and
the absolute value of the fundamental gap $E_0$ decreases (see
Fig. 2), the spin $g^*_0$ value at the band edge decreases. On the
other hand, with increasing T the electrons populate higher LLs
and band's nonparabolicity comes more and more into play. The
latter is known to make the $g$ value less negative (see Refs.
[10, 24], and Fig. 3). Thus, as $T$ increases, the average $g^*$
decreases or increases depending on the relative strength of the
two effects. It is now clear why putting into calculations only
the dilatational part of the gap variation makes the first effect
weaker, i.e. it favors increase of the $g$ values. We emphasize
that we do not use in our calculations Eq. (7), it is quoted only
to make clear the dependence of $g^*_0$ on $E_0$. Calculating the
electron energies away from the band edge we deal with the effects
of band's nonparabolicity and inversion asymmetry, as explained
above. We use the following band parameters of GaAs at $T$ = 0:
$E_{P_0}$ = 27.865eV, $E_{P_1}$ = 2.361eV, $E_0$ = -1.519 eV,
$G_0$ = -1.86 eV, $E_1$ = 2.969 eV, $G_1$ = 3.14 eV, $C$ =
-2.3107, $C'$ = -0.0375. The zero of energy is chosen at the
conduction band edge. The above values are the same as those
established by an overall fit of various experiments on GaAs (see
Ref. [10]) with the exception of the far-band contribution $C'$,
as discussed below. It is assumed that only $E_0$ depends on the
temperature, see discussion below.

In Fig. 3 we plot the $g$ values for consecutive LLs at two
temperatures, as determined from definition (3), in which the
energies are calculated for $B$ = 4 T using the procedure
described above. It is seen that, for a given temperature, as the
energy increases with the growing Landau number $n$, the $g$
factor increases due to band's nonparabolicity. On the other hand,
for a given LL the $g$ factor decreases with increasing
temperature. The number of points for a given temperature
indicates how many LLs are involved in the averaging, for still
higher LLs the occupation by electrons is so small that their
contribution is negligible. It should be noted that the $g$
factors shown in Fig. 3 do not saturate at high energies at the
free electron value of +2, as predicted by simple versions of the
three-level or five-level {\kp} models (see e.g. [24]). The reason
is that our present theory includes the bulk inversion asymmetry
(also called the Dresselhaus effect), which is manifested by the
appearance of the matrix element Q. Also, the increase of $g$
factor with the energy, as shown in Fig. 3, is stronger than
linear, whereas the one shown in Fig. 8 of Ref. [10] is weaker
than linear. This is caused by the fact that in Fig. 3 the energy
increases by going to high LLs (at constant $B$), whereas in Ref.
[10] the energy is increased by going to high magnetic fields (at
$n$ = 0).
\section{\label{sec:level1}RESULTS AND DISCUSSION\protect\\ \lowercase{}}

Fig. 4 shows the temperature dependent $g$ values of electrons
extracted from time-resolved Faraday rotation measurements
performed at $B$ = 4 T on undoped bulk GaAs (full squares). In
addition, data taken from the literature and our calculations are
shown. As to the experiments, the previously published data
obtained with the use of time-resolved photoluminescence [1-3] and
time-resolved Kerr rotation [25] are compared with our present
results obtained by the time-resolved Faraday rotation. All these
time-resolved experiments measure the phase coherence of spin
states excited by circularly polarized light in the Voigt geometry
and differ mostly in its detection. It is seen that the obtained
data agree well with each other and all show an almost linear
increase of $g$ with the temperature. However, there is some
discrepancy between various measured $g$ values at $T$ = 0, cf.
[23, 26, 27]. Since all the $g$ values presented in Fig. 4 were
measured at magnetic fields $B$ of a few Tesla, they should be
\emph{higher} than the band edge value $g^*_0$ since they are
affected by band's nonparabolicity. In this situation we take the
value of the far-band contributions $C'$, as given above, to give
our measured value of $g^*$ = -0.452 at $T$ = 0 and $B$ = 4 T.
This corresponds to the band edge value $g^*_0$ = -0.472.
Preference to a different experimental band edge $g$ value would
simply require a slightly different $C'$.

As far as the theory is concerned, we show two calculations. The
dashed line indicates an average value of $g$ computed with the
help of Eqs. (4), (5) and (6), in which the $E_n^{\pm}$ energies
are calculated for $B$ = 4 T using the procedure outlined above.
It is seen that, as the temperature increases and more LLs become
populated, the band nonparabolicity effect is stronger than the
dilatational decrease of the $E_0$ gap, and the average
$\overline{g^*}(T)$ increases. Still, the theoretical increase
does not quite follow the experimental values for $T>$ 60 K. The
reason is that our simplified formula (2) assumes the spin
splitting to be independent of $k_z$. This means that, as we
average over the energy, the $g$ value "jumps" from one LL to the
next, as shown in Fig. 3. In reality, however, the $g$ value
depends also on $k_z$. As follows from the simplified three-level
{\kp} model, the spin $g$ value depends in fact on $k_z$
\emph{similarly} to its dependence on the orbital energy $\hbar
\omega_c(n+1/2)$. In other words, the decisive quantity is the
total electron energy (see [8, 24, 28]). For the second
calculation we \emph{assume} that this is the case also in the
complete five-level {\kp} model and we interpolate the $g$ value
to vary linearly with the energy (related to $k_z$) between
consecutive Landau levels. If this is done, the averaging
procedure gives the $g$ values indicated in Fig. 4 by the solid
line. It is seen that now the theory is in an excellent agreement
with the experimental data.
\begin{figure}
\includegraphics[scale=0.92,angle=0, bb = 20 25 291 229]{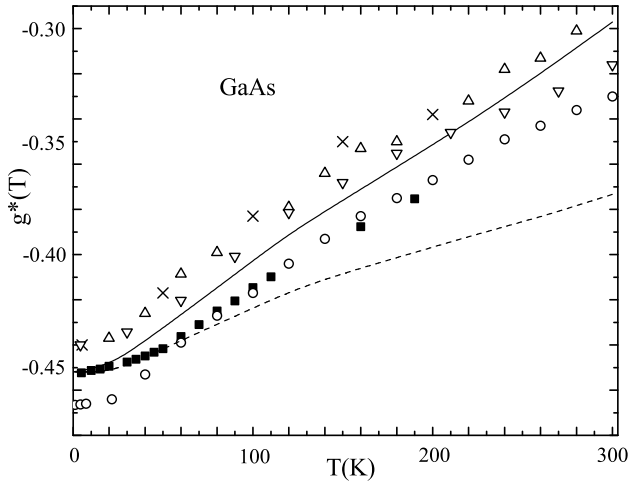}
\caption{\label{fig:epsart}{Measured and calculated spin $g$ values of electrons in GaAs versus temperature.
Experimental data: full squares - our results, crosses - Ref. [1], triangles - Ref [2], circles - Ref. [3],
reversed triangles - Ref. [25]. Theory: dashed line - calculation neglecting $k_z$ - dependence of $g^*_n$;
solid line - calculation assuming interpolated $k_z$ - dependence of $g^*_n$, see text.}} \label{fig4th}
\end{figure}
The moral of the story is that the almost linear increase of the average spin $g$ value with the temperature in
GaAs is caused by the dominating effect of band's nonparabolicity over the dilatational decrease of the
fundamental gap. It should be emphasized that both our experimental data as well as our two theoretical curves
approach the absolute zero of temperature at almost vanishing slope. It is important that the improved
experiment and theory exhibit this property.

Coming to the discussion of our results, one should keep in mind that our theoretical description, although
representing a considerable improvement over the procedure of Ref. [4], is still approximate. The reason is that
we made an assumption about the $k_z$ dependence of the $g$ factor. This assumption is reasonable and it allows
us to calculate the ${\cal E}_n^{\pm}$ energies and carry out the averaging procedure over LLs in a relatively
simple way. Still, in principle one should include the $k_z$ terms from the beginning in the computation
procedure, calculate the $k_z$ dependence of different LLs and carry out a numerical summation over $k_z$
values. This, however, makes the problem difficult for two reasons. First, if the $k_z$ terms are included, one
must take much larger truncated determinants to obtain convergent values of ${\cal E}_{n}^{\pm}(k_z)$ energies.
Second, when performing the averaging procedure one must integrate numerically over $k_z$ for each LL separately
since in principle each LL has a different $k_z$ dependence. Our simplifying assumption allowed us to get around
the above difficulties.

As to the work of other authors, Oestreich and Ruhle [1] put into
the theoretical description the total temperature change of the
energy gap but neglected band's nonparabolicity. Such a
description predicted a decrease of the $g$ factor with
temperature which contradicted the experimental results. Oestreich
et al [2] and Huebner et al [3] included in their description
band's nonparabolicity but neglected important features of the
band structure: the interband matrix element of momentum Q and of
the spin orbit interaction $\overline{\Delta}$ between the
($\Gamma^v_7$, $\Gamma^v_8$) and ($\Gamma^c_7$, $\Gamma^c_8$)
multiplets. Both these quantities have an important influence on
the $g$ value, which we illustrate numerically. Using our band
parameters and $\overline{\Delta}\neq 0$ the band edge $g$ value
is $g^*_0$ = -0.472, while for the same parameters and
$\overline{\Delta}$ = 0 there is $g^*_0$ = -0.346 (see Eq. (7)).
The element Q does not enter the conduction band edge quantities,
but at $T$ = 300 K and $B$ = 4 T our calculated average value of
$g$ including Q is $\overline{g^*}(T)$ = -0.297, while for Q = 0
the calculated average is $\overline{g^*}$ = -0.396. Also, Refs.
[2], [3] did not include the Landau quantization of the conduction
band, so that the calculated average $g$ value did not depend on
the magnetic field intensity. Finally, and this is the decisive
point, it was assumed that the $g$ value depends on the
temperature via the total (i.e. measured) change of the
fundamental gap, whereas one should take the change due to thermal
lattice dilatation alone. When this is done, there is no need to
introduce a temperature variation of the interband matrix element
of momentum $P_0$, as was done in Ref. [3]. Consideration of the
momentum operator and of the wave functions indicates that the
interband momentum matrix element should also scale with the
lattice dilatation, its square should change no more than 0.1 $\%$
in the temperature range from 0 to 300 K, so it can be neglected
[4].

The latest analysis by Litvinenko et al [4], including the matrix
elements $\overline{\Delta}$ and Q and the band nonparabolicity on
one side and taking only the dilatation contribution to the change
of the energy gap on the other, was able to account for the first
time for the experimental increase of the spin $g$ factor with the
temperature. However, this analysis assumed no $k_z$ dependence of
the Landau levels and, consistently, took the average electron
energy for the Boltzmann statistics to be equal to $kT$. This
corresponds roughly to our dashed curve in Fig. 4, which assumes
no dependence of $g$ on $k_z$. If one includes the $k_z$
dependence of the energy, the average nondegenerate electron
energy is $\overline{\cal E} = (3/2)kT$ which, upon using Fig. 3
for $g_n(\cal {E})$, corresponds quite well to the solid line in
Fig. 4. In other words, if one applies a proper averaging
procedure to the $g$ value, the resulting $\overline{g^*}(T)$
corresponds quite well to averaging the electron energy
$\overline{\cal E}$ and taking $g^*(\overline{\cal E})$. One
should add that, taking the thermal change of gap due to the
dilatation alone, Litvinenko et al [4] were able to account
correctly for the temperature dependence of the electron spin
splitting in InSb. All this, together with the published analysis
of $m^*(T)$ in GaAs [14, 15], strongly confirms that both orbital
and spin quantizations of the electron spectrum in a magnetic
field are governed by the change of the energy gap due to lattice
dilatation alone.

\section{\label{sec:level1}SUMMARY\protect\\ \lowercase{}}

We investigated experimentally and theoretically the controversial
problem of temperature dependence of the spin $g$ factor in GaAs.
The time-resolved Faraday rotation technique was used to show
that, in agreement with other data, the $g$ factor increases
almost linearly with $T$ up to room temperature. This increase is
successfully described by the five-level {\kp} model of the band
structure for GaAs. As the temperature increases there occur two
effects having an opposite influence on the $g$ value. On one
hand, a decrease of the fundamental energy gap caused by the
thermal lattice dilatation leads to a decrease of $g$. On the
other hand, at higher temperatures electrons populate higher
Landau levels which, due to band's nonparabolicity, leads to an
increase of the average $\overline{g}(T)$ value. A very good
agreement of our theory with the experimental increase of
$\overline{g}(T)$ indicates that the band nonparabolicity effect
dominates. Our interpretation confirms the validity of the
complete five-level {\kp} model for the conduction band of GaAs.

\begin{acknowledgments} W. Z. and P. P. acknowledge a financial
support of Polish Ministry of Science and Higher Education through
 Laboratory of Physical Foundations of Information Processing.
\end{acknowledgments}


\begin{thebibliography}
{99}\label{sec:TeXbooks}

\bibitem{pp1} M. Oestreich and W. W. Ruhle, Phys. Rev. Lett. \textbf{74}, 2315
(1995).

\bibitem{pp2} M. Oestreich, S. Hallstein, A. P. Heberle, K. Eberl, E. Bauser, and W. W. Ruhle, Phys. Rev.
B \textbf{53}, 7911 (1996).

\bibitem{pp3} J. Huebner, S. Dohrmann, D. Hagele, and M. Oestreich, \emph{Preprint} arXiv: cond-mat/0608534v4 (2006).

\bibitem{pp4} K. L. Litvinenko, L. Nikzad, C. R. Pidgeon, J. Allam, L. F. Cohen, T. Ashley, M. Emeny, W.
Zawadzki, and B. N. Murdin, Phys. Rev. B \textbf{77}, 033204 (2008).

\bibitem{pp5} D. D. Awschalom, D. Loss, N. Samarth, \emph{Semiconductor Spintronics and Quantum Computation}
(Springer-Verlag, Heidelberg, 2002).

\bibitem{pp6} R. Bratschitsch, Z. Chen, S. T. Cundiff, E. A. Zhukov, D. R. Yakovlev, M. Bayer, G. Karczewski,
T. Wojtowicz, and J. Kossut, Appl. Phys. Lett. \textbf{89}, 221113 (2006).

\bibitem{pp7} Z. Chen, S. Carter, R. Bratschitsch, P. Dawson, and S. T. Cundiff, Nature Physics \textbf{3}, 265 (2007).

\bibitem{pp8} R. Bowers and Y. Yafet, Phys. Rev. {\bf 115}, 1165 (1959).

\bibitem{pp9} C. R. Pidgeon and R. N. Brown, Phys. Rev. \textbf{146}, 575 (1966).

\bibitem{pp10} P. Pfeffer and W. Zawadzki, Phys. Rev. B \textbf{41}, 1561
(1990).

\bibitem{pp11} P. Pfeffer and W. Zawadzki, Phys. Rev. B \textbf{53}, 12813 (1996).

\bibitem{pp12} P. Pfeffer and W. Zawadzki, Phys. Rev. B \textbf{74}, 115309
(2006); \textbf{74}, 233303 (2006).

\bibitem{pp13} H. Ehrenreich, J. Phys. Chem. Solids \textbf{2}, 131 (1959).

\bibitem{pp14} R. A. Stradling and R. A. Wood, J. Phys. C \textbf{3}, L94 (1970).

\bibitem{pp15} H. Hazama, T. Sugimasa, T. Imachi, and C. Hamaguchi, J. Phys. Soc. Jpn. \textbf{55}, 1282 (1986).

\bibitem{pp16} S. A. Lourenco, I. F. L. Dias, J. L. Duarte, E. Laureto, L. C. Pocas, D. O. Toginho Filho,
and J. R. Leite, Brazilian Journ. Phys. \textbf{34}, 517 (2004).

\bibitem{pp17} P. Lautenschlager, M. Garriga, S. Logothetidis, and M. Cardona, Phys. Rev. B \textbf{35}, 9174
(1987).

\bibitem{pp18} S. I. Novikova, Fiz. Tverd. Tela \textbf{3} 178 (1961),
(Sov. Phys. - Solid St. \textbf{3}, 129 (1961)).

\bibitem{pp19} T. Soma, J. Satoh, and H. Matsuo, Solid St. Commun. \textbf{42}, 889 (1982).

\bibitem{pp20} E. Grilli, M. Guzzi, R. Zamboni, and L. Pavesi, Phys. Rev. B \textbf{45}, 1638 (1992).

\bibitem{pp21} R. Paessler, phys. stat. sol. (b) \textbf{200}, 155 (1997).

\bibitem{pp22} F. H. Pollak, C. W. Higginbotham, M. Cardona, J. Phys. Soc. Jpn. Suppl. \textbf{21}, 20 (1966).

\bibitem{pp23} C. Hermann and C. Weisbuch, Phys. Rev. B \textbf{15}, 823 (1977).

\bibitem{pp24} W. Zawadzki, Physics Lett. {\bf 4}, 190 (1963).

\bibitem{pp25} P. E. Hohage, G. Bacher, D. Reuter and A. D. Wieck, Appl. Phys. Lett. {\bf 89}, 231101 (2006).

\bibitem{pp26} A. P. Heberle, W. W. Ruhle, and K. Ploog, Phys. Rev. Lett. \textbf{72}, 3887
(1994).

\bibitem{pp27} M. Krapf, G. Denninger, H. Pascher, G. Weimann, and W. Schlapp, Solid St. Commun.
\textbf{74}, 1141 (1990).

\bibitem{pp28} W. Zawadzki, in \emph{Narrow Gap Semiconductors, Physics and Applications}, edited by W. Zawadzki,
(Springer, Berlin, 1980), p. 85.



\end{thebibliography}
\end{document}